\documentclass[useAMS,usenatbib,usegraphicx]{mn2e}
\usepackage{aas_macros}
\usepackage{url}
\usepackage{color}
\usepackage{array}
\usepackage{amsmath,amssymb,mathrsfs}

\begin{document}

  \title[Accelerating pulsar timing data analysis]{Accelerating pulsar timing data
    analysis}
  
  \author[van Haasteren]{
Rutger~van Haasteren$^{1}$\footnotemark
  \\
    $^1$Max-Planck-Institut f\"ur Gravitationsphysik (Albert-Einstein-Institut),
      D-30167 Hannover, Germany}
 
  \date{printed \today}

  \maketitle

  \begin{abstract}
    The analysis of pulsar timing data, especially in pulsar timing array (PTA)
    projects, has encountered practical difficulties:
    evaluating the likelihood and/or correlation-based statistics can become
    prohibitively computationally expensive for large datasets. In situations
    where a stochastic signal of interest has a power spectral density that
    dominates the noise in a limited bandwidth of the total frequency domain
    (e.g. the isotropic background of gravitational waves), a linear
    transformation exists that transforms the timing residuals to a basis in
    which virtually all the information about the stochastic signal of interest
    is contained in a small fraction of basis vectors. By only considering such
    a small subset of these ``generalised residuals'', the
    dimensionality of the data analysis problem is greatly reduced, which can
    cause a large speedup in the evaluation of the likelihood: the ABC-method
    (Acceleration By Compression).
    The compression fidelity, calculable with crude estimates of the signal and
    noise, can be used to determine how far a dataset can be compressed without
    significant loss of information. Both direct tests on the likelihood, and
    Bayesian analysis of mock data, show that the signal can be recovered as
    well as with an analysis of uncompressed data. In the analysis of IPTA Mock Data
    Challenge datasets, speedups of a factor of three orders of
    magnitude are demonstrated. For realistic PTA datasets the acceleration may
    become greater than six orders of magnitude due to the low signal to noise
    ratio.
  \end{abstract}

  \begin{keywords}
    gravitational waves -- pulsars: general -- methods: data analysis
  \end{keywords}

\footnotetext{Email: vhaasteren@gmail.com}

  \section{Introduction}
    In the past several decades, pulsar timing has been successfully used to
    study a wide range of science. Past successes include the confirmation of
    gravitational waves \citep{Taylor1982}, and very accurate tests of general
    relativity \citep{Kramer2006}. The interesting science of these examples stems
    from the fact that accurate measurements of the times of arrival (TOAs) of
    the radio pulses allow for a precise determination of the trajectory of
    the pulsar relative to the Earth. This is possible because the TOAs can be
    accurately accounted for by current models of the pulsar trajectory,
    pulse propagation, and pulsar spin evolution in relativistic gravity. 

    Among on-going pulsar timing projects are Pulsar Timing Arrays (PTAs), which
    are programmes designed to detect low-frequency ($10^{-9}$---$10^{-8}$Hz)
    extragalactic gravitational-waves (GWs) directly, by using a set of Galactic
    millisecond pulsars as nearly-perfect Einstein clocks \citep{Foster1990}.
    GWs perturb space-time between the pulsars and the Earth, and this creates
    detectable deviations from the strict periodicity in the TOAs
    \citep{Estabrook1975, Sazhin1978, Detweiler1979}. One of
    the main source candidates for PTAs is an isotropic stochastic background of
    gravitational waves (GWB), thought to be generated by a large number of
    massive black-hole binaries located at the centres of galaxies
    \citep{Begelman1980, Phinney2001, Jaffe2003, Wyithe2003, Sesana2008},
    by relic gravitational-waves \citep{Grishchuk2005}, or, more
    speculatively, by oscillating cosmic-string loops \citep{Damour2005,
    Olmez2010, Sanidas2012}.

    The analysis of pulsar timing data, and even more so PTA data, can become
    prohibitively time-consuming for large datasets.  This is especially true
    for Bayesian data analysis methods, like the analysis of PTA data
    \citep[][hereafter vHLML]{vanhaasteren2009}, and the
    correction for dispersion measure
    variations (Lee et al., in prep.).
    Typically, the computational cost scales as $n^3$ or $n^2$, with $n$ the
    total number of observations; the computational difficulties will
    increase sharply over time.

    In this work, one possible solution for the computational difficulties is
    explored in the case the signal of interest is a time-correlated stochastic
    signal: the ABC-method (Acceleration By Compression). The
    ABC-method is based on
    lossy linear data compression. By significantly reducing the
    dimensionality of the problem, the evaluation of computationally expensive
    quantities can be greatly accelerated. We specifically focus on the
    International PTA (IPTA) Mock Data Challenge (released by M.~Keith,
    K.~J.~Lee, and F.~A.~Jenet\footnote{
    \url{http://www.ipta4gw.org/?page_id=214}
}), in which the GWB is a good example of a compressible stochastic signal.

    The outline of the paper is as follows. In Section~\ref{sec:ptaan} we
    briefly review the relevant theory of pulsar timing observations, with a
    special attention to the likelihood in the presence of time-correlated
    stochastic signals. We introduce the ABC-method, and the compressibility of
    datasets, in
    Section~\ref{sec:lossycompression}. In Section~\ref{sec:practice} we look
    into some of the practicalities concerned with compression of PTA data, and
    investigate the computational demand of different terms in the evaluation of
    the likelihood. In that section, we provide and test a method based on cubic
    spline interpolation to estimate the compressed covariance matrix. This
    causes an extra speedup of a few orders of magnitude.  Finally we present
    our conclusions in Section~\ref{sec:conclusions}.

  \section{PTA data analysis} \label{sec:ptaan}
    The typical data processing pipeline for pulsar timing observations
    processes the raw baseband data in several data reduction steps, where at
    each data reduction step the data volume is drastically reduced. The data
    reduction steps condense the scientifically interesting information into a
    significantly smaller number of data points, sometimes
    mitigating noise in the process. At the end of the pipeline we are left with
    TOAs.

    This work proposes a method to compress the TOA data even further to what we
    call generalised residuals. The data compression is based on
    the likelihood for the TOAs and the Fisher information, with information
    preserved only for a specific stochastic signal. To this end, we review the
    theory of TOAs, the likelihood, and inclusion of the timing-model in this
    section.

    \subsection{The likelihood}
      We consider $k$ pulsars, with $n^{\prime}_{a}$ TOAs for the $a$-th pulsar,
      where the $n^{\prime}=\sum_{a=1}^{k}n^{\prime}_{a}$ TOAs are described as
      an addition of a
      deterministic and a stochastic part. In the observations this distinction
      is blurred because we cannot
      fully separate the stochastic contributions from the deterministic
      contributions. In practice we therefore work with timing
      residuals that are produced using first estimates $\beta_{0i}$ of the
      $m$ timing-model parameters $\beta_i$ ($i$ between $1$ and $m$); this
      initial guess is usually assumed to be accurate enough to use a linear
      approximation of the
      timing-model \citep{Edwards2006}. Here $m=\sum_{a=1}^{k}m_{a}$ is the sum
      of the number of timing-model parameters of all the individual pulsars.
      In this linear approximation, the
      timing-residuals depend on $\xi_i = \beta_i - \beta_{0i}$ as:
      \begin{equation}
	\vec{\delta t}^{\prime} = \vec{\delta t}^{\rm prf} + M\vec{\xi},
	\label{eq:prefitresidual}
      \end{equation}
      where $\vec{\delta t}^{\prime}$ are the timing-residuals in the linear
      approximation to the timing-model, $\vec{\delta t}^{\rm prf}$ is the
      vector of pre-fit timing-residuals, $\vec{\xi}$ is the vector with
      timing-model parameters for all $k$ pulsars, and the $(n^{\prime}\times m)$ matrix
      $M$ is the so-called design matrix
      \citep[see e.g. \S $15.4$ of][vHLML]{Press1992}, which describes how the
      timing-residuals depend on the model parameters. As an example, for a
      simple timing model which only contains quadratic spindown, the matrix $M$
      is a $(n^{\prime}\times 3)$ matrix, with the $j$-th column describing a
      $(j-1)$-th order polynomial. The elements of $M$ are then:
      $t_{i}^{j-1}$, with $t_{i}$ the $i$-th TOA.

      Identical to vHLML and \citet[][hereafter vHL]{vanhaasteren2012}, we
      model the stochastic contributions to the TOAs as a time-correlated
      stochastic signal, described by a
      random Gaussian process.  The corresponding likelihood is equal to:
      \begin{equation}
	\label{eq:likelihood}
	P\left(\vec{\delta t}^{\prime} | \vec{\xi}, \vec{\phi}\right) =
	\frac{
	  \exp\left[-\frac{1}{2} \left(\vec{\delta
	    t}^{\prime} - M\vec{\xi}\right)^{T}C^{\prime-1}\left(\vec{\delta 
	    t}^{\prime} - M\vec{\xi}\right)\right]
	}{
	  \sqrt{(2\pi)^{n^{\prime}}\det C^{\prime}}
	},
      \end{equation}
      where $\vec{\phi}$ is the vector describing all the stochastic model
      parameters, and $C^\prime = C^\prime(\vec{\phi})$ is the covariance matrix
      of the sum of all stochastic signals. This includes the measurement
      uncertainties, the timing noise (red spin noise), and a possible GWB.
    
    \subsection{Marginalising over the timing-model} \label{sec:martm}
      Using Equation~(\ref{eq:likelihood}) is computationally not very
      efficient because of the large number of timing model parameters.
      However, in the case of uniform priors (vHLML) and
      Gaussian priors (vHL) it is possible to analytically marginalise the
      posterior distribution over the timing model parameters.
      In the remainder of this work we assume no prior information about the
      timing model parameters, and use uniform priors.

      In their search for a simplified representation of the analytic
      marginalisation procedure, vHL decomposed the design matrix into an
      orthogonal basis based on the singular value decomposition 
      $M = U \Sigma V^{*}$, where $U$ and $V$ are $(n^{\prime}\times n^{\prime})$ and $(m\times
      m)$ orthogonal matrices, and $\Sigma$ is an $(n^{\prime}\times m)$ diagonal matrix.
      The first $m$ columns of $U$ span the column space of $M$, and the last
      $n=n^{\prime}-m$ columns of $U$ span the complement. We denote these two subspace
      bases as $F$ and $G$ respectively: $U = \begin{array}{cc}(F & G)\end{array}$. In
      Section~\ref{sec:marislos} we show that $G$ is actually a lossless data
      compression matrix.

      Now, integrating over $\vec{\xi}$, our
      marginalised likelihood becomes (vHLML):
      \begin{eqnarray}
	\int \! \mathrm{d}^{m}\vec{\xi} P(\vec{\delta t}^{\prime} | \vec{\xi},
	\vec{\phi}) &=&
	\frac{\sqrt{\det \left(F^TC^{\prime-1}F\right)^{-1}}}{\sqrt{(2\pi)^{n}\det
	C^\prime}} \times \nonumber \\
	& & \exp\left(-\frac{1}{2} \vec{\delta
	t}^{\prime T}C_{P}^{-1}\vec{\delta t}^{\prime}\right),
	\label{eq:marginalisedlikelihood}
      \end{eqnarray}
      with:
      \begin{eqnarray}
	\label{eq:cprime}
	C_{P}^{-1} &=& C^{\prime -1} - C^{\prime -1}F\left(F^{T} C^{\prime
	-1}F\right)^{-1}F^{T} C^{\prime -1} \\
	C_{P} &=& GG^{T}C^{\prime}GG^{T}, \nonumber
      \end{eqnarray}
      where the singular matrix $C_{P}^{-1}$ is the inverse of $C_{P}$ in the
      non-singular subspace of its basis.
      The singular matrix matrix $C_{P}$
      is the post-fit covariance matrix of the timing-residuals
      \citep[vHL; ][hereafter D12]{Demorest2012}.
      D12 use a pseudo-inverse based
      on a singular value decomposition of $C^{\prime}$ to evaluate $C_{P}^{-1}$
      in their evaluation of a GWB detection statistic; this is equivalent to
      marginalising over the timing model parameters (vHL).

  \section{The ABC-method} \label{sec:lossycompression}
    Data compression is the encoding of information in a smaller data volume
    than the original information data volume. This can be done without losing
    information (lossless), or with losing information (lossy) \citep{Wade1994}.
    We would like to use data compression to reduce our data volume, with the
    aim of speeding up the computations that are necessary for the analysis of
    PTA data. In this work we compress the data in such a way to retain the
    sensitivity to one stochastic signal (e.g. the isotropic background of
    gravitational waves): the ``ABC-method'' (Acceleration By Compression).

    In Section~\ref{sec:marislos} we show that marginalisation over the timing
    model parameters is equivalent to lossless data compression. In
    Section~\ref{sec:lincom} we expand the data compression formalism, and show
    how to construct a basis in which sensitivity to a particular signal is
    retained. We define the corresponding compression fidelity in
    Section~\ref{sec:comfid}. Finally, in Section~\ref{sec:interpret} and
    Section~\ref{sec:compressibility}, we discuss how to interpret the compressed
    basis of generalised residuals, and how far a dataset can be compressed
    without significant loss of information.

    \subsection{Marginalisation = lossless data compression} \label{sec:marislos}
      vHL showed that Equation~(\ref{eq:marginalisedlikelihood}) can be
      rewritten as:
      \begin{equation}
	\label{eq:marginalisedlikelihoodnew}
	\int \! \mathrm{d}^{m}\vec{\xi} P(\vec{\delta t}^{\prime} | \vec{\xi}, \vec{\phi})
	= \frac{
	 \exp\left[-\frac{1}{2} \vec{\delta
	 t}^{\prime T}G \left(G^{T}C^{\prime}G\right)^{-1}G^{T}
	 \vec{\delta t}^{\prime}\right]
	}{\sqrt{(2\pi)^{n}\det
	\left(G^{T}C^{\prime}G\right)}},
      \end{equation}
      with notation as in Section~\ref{sec:martm}.  This is an unmarginalised
      likelihood of a random Gaussian process in $n$ dimensions, with data
      $\vec{\delta t}=G^{T}\vec{\delta t}^{\prime}$ and covariance matrix $C =
      G^{T}C^{\prime}G$. The dimensionality of the data is reduced from
      $n^{\prime}$ to $n$ due to the marginalisation process.  From here
      onwards, we start the convention that a prime superscript denotes that a
      vector or a covariance matrix lies in in the larger unmarginalised space,
      whereas no prime denotes that either of them lies in the
      marginalised space.  The vector $\vec{\delta t}$ contains all the
      information about all stochastic signals: marginalisation over the timing
      model parameters is the same as lossless linear data compression in this
      formalism.  The matrix $G$ is our linear data compression matrix, and
      $\vec{\delta t}$ is our vector of reduced data.

    \subsection{Lossy linear data compression} \label{sec:lincom}
      We would like to compress the reduced data $\vec{\delta t}$ even
      further, without losing too much information about the stochastic signal
      of our interest.  We expect this to be possible, since usually the signal
      and the noise differ in power spectral density. Only some parts of the
      spectrum are dominated by the signal; other parts are dominated by the
      noise. The data compression scheme in this work is based on throwing away
      the parts of the data that are dominated by the noise by using linear data
      compression: $\vec{x} = H^{T}\vec{\delta t}$, with $\vec{x}$ the
      compressed data, or ``generalised residuals'' as we will call them, and
      $H$ the \emph{compression matrix}.  Here the number of columns of $H$ is
      less than the number of rows, where
      we define the compression to be the total number of timing residuals
      divided by the number of compressed generalised timing residuals.
      We derive one possible scheme to construct a suitable $H$ in this section.

      In order to determine how much information about our signal of interest is
      in our data, we use the Fisher information. We acknowledge that formally
      the Fisher information does not completely quantify how well a parameter
      can be confined with a specific dataset, especially in the case of a low
      signal-to-noise ratio \citep[e.g.][]{Vallisneri2008}, but in this
      exploratory work we consider the Fisher information as a sufficient first
      attempt. Denoting the log-likelihood of
      Equation~(\ref{eq:marginalisedlikelihoodnew}) as $\Lambda$, we find for
      the Fisher information:
      \begin{equation}
        \label{eq:fisherinformation}
	\mathcal{I}_{\theta\phi} =
	\left\langle-\frac{\partial^2\Lambda}{\partial \theta\partial
	\phi}\right\rangle =
	\frac{1}{2}\text{Tr}\left(
	    \frac{\partial C}{\partial \theta}C^{-1}
	    \frac{\partial C}{\partial \phi}C^{-1}
	    \right),
      \end{equation}
      where $\mathcal{I}_{\theta\phi}$ is the Fisher information, and $\phi$ and
      $\theta$ are model parameters that affect the signal power spectral density.
      Suppose that the stochastic processes in the reduced data $\vec{\delta t}$
      are described by the covariance matrix $C =  \Sigma +
      a^{2}S$, where $\Sigma$ is the covariance matrix of the noise, and $S$ is
      the covariance matrix of the signal of interest with amplitude $a^2$. We
      would like to know which basis vectors have the largest contribution to
      the Fisher information, which would be easiest to determine if we could
      completely diagonalise the matrices in the trace of
      Equation~\ref{eq:fisherinformation}. This is possible with a
      non-orthogonal transformation. Even though the inner product is not
      preserved in such a transformation, the trace remains invariant. We use
      a square root of the noise matrix, $\Sigma_{w}^{-1/2}=\Sigma^{-1/2}$, to
      do this. For the moment we assume that this estimate of $\Sigma_{w}$ is
      indeed correct, but in Section~\ref{sec:interpret} we argue that an
      inaccurate noise estimate still results in a usable compression.
      In this new basis, the whitened data and covariance become
      $\vec{{\delta t}^{\rm w}} = \Sigma_w^{-1/2}\vec{\delta t}$ and
      $C^{\rm w}=\Sigma_w^{-1/2}C\Sigma_w^{-1/2}$. The maximum sensitivity
      based on the Fisher information now
      has a simple form:
      \begin{equation}
      \label{eq:sensitivity}
	\frac{a^2}{\text{Var}(a)} \leq
	a^{2}\mathcal{I}_{aa} =
	2\sum_{i=1}^{n}\frac{a^2\lambda_i^2}{(1+a\lambda_i)^2},
      \end{equation}
      where $\lambda_i$ is the $i$-th eigenvalue of $C^{\rm w}$. The $a\lambda_i$
      should be interpreted as signal to noise ratios.
      We can only
      evaluate Equation~(\ref{eq:sensitivity}) if we have complete knowledge
      of the signal $S$, the signal amplitude $a$, and the noise $\Sigma$.
      However, the $\lambda_i$ and the corresponding basis vectors do not depend
      on $a$, which means we can examine the sensitivity to $a$ as a function of
      the number of $\lambda_i$ we include. Here, we do assume knowledge of
      $\Sigma$ and $S$.

      In the limit where $a$ is large, the strong signal limit, we can neglect
      the one in the denominator of the sum of Equation~(\ref{eq:sensitivity}),
      which makes all terms in the sum equal.
      This means that all generalised residuals carry equal
      information as is expected in such a case:
      the noise is negligible compared to
      the signal, so no parts of the signal are buried under the noise. In the
      strong signal limit, data compression is therefore not possible.  Note
      that the sensitivity is then proportional to the number of generalised
      residuals, as it should.

      In the
      limit that $a$ is small, the low-signal limit, we can neglect all terms
      $a\lambda_i$ in the denominator of the sum, making the sensitivity equal
      to the sum of all $a^2\lambda_i^2$. The distribution of values of the
      $\lambda_i$ eigenvalues is determined by the power spectral density of the
      signal compared to the noise. If the signal spectrum is the same as the
      noise spectrum, all the $\lambda_i$ will be identical.  However, if the
      signal has a different spectrum than the noise, the $\lambda_i$ can span a
      wide range of values, where the large $\lambda_i$ correspond to basis
      vectors where the signal is relatively large compared to the noise. In
      this case there are nearly-redundant data points, and compression is
      possible.
      
    \subsection{The compression fidelity} \label{sec:comfid}
      We define
      the fidelity $\mathscr{F} \in [0,1]$ to be the fraction of the total
      sensitivity we retain in our compressed data.  We choose the number of
      generalised residuals that we keep, $l$, to be the smallest number such
      that:
      \begin{equation}
        \label{eq:fidelity}
        \frac{\sum_{i=1}^{l}
	  \lambda_i^2 / \left(1+a\lambda_i \right)^2}{\sum_{i=1}^{n}\lambda_i^2 / \left(1+a\lambda_i \right)^2} \geq \mathscr{F},
      \end{equation}
      where we have ordered the $\lambda_i$ to have the largest values for the
      lowest indices. We typically work with $\mathscr{F} \geq 0.99$, which in
      favourable cases like the IPTA Mock Data Challenge allows for compressions
      greater than $10$: less than $10\%$ of the original data volume is kept.

      Computationally, we suggest to use a singular value decomposition to
      produce the eigenvalues and eigenvectors of $C^{\rm w}$, where our
      fidelity criterion, Equation~(\ref{eq:fidelity}), keeps only $l$ of the
      $n$ generalised residuals ${\delta t}^{\rm w}_i$.
      We construct the $(n \times l)$ matrix $W$ as
      consisting of the columns
      of the $l$ eigenvectors that belong to the selected eigenvalues. The data
      compression matrix is now: $H = \Sigma_w^{-1/2}W$.
      Using
      Equation~(\ref{eq:marginalisedlikelihoodnew}), we now find for the
      likelihood of the compressed data $\vec{x}=H^{T}\vec{\delta t}$:
      \begin{equation}
        \label{eq:compressedlikelihood}
	P(\vec{x} | \vec{\phi})
	= \frac{
	 \exp\left[-\frac{1}{2} \vec{x}^{T} \left(H^{T}CH\right)^{-1}
	\vec{x}\right]
	}{\sqrt{(2\pi)^{l}\det\left(\Sigma_w \right)\det
	\left(H^{T}CH\right)}},
      \end{equation}
      where the extra determinant of $\Sigma_w$ comes from the whitening, and
      can be ignored in practice as it is absorbed in the overall normalisation
      constant. This equation is the basis of the ABC-method, as the
      computationally expensive inversion has been replaced with a
      lower-dimensional one.

      Equation~(\ref{eq:compressedlikelihood}) is completely general, and can
      readily be applied to realistic datasets. As in \citet{vanhaasteren2011},
      all timing-model parameters and jumps can be included in the likelihood,
      and are therefore by design part of the data compression scheme. We
      therefore expect not to encounter any difficulties in applying
      data compression to realistic data sets, even though in this work we only
      test the effectiveness on the Mock Data Challenge.

    \subsection{Interpreting the compressed basis} \label{sec:interpret}
      As we have discussed in Section~\ref{sec:marislos}, marginalising over the
      timing model is the same as linear data compressing to the subspace of the
      original data $\vec{\delta t}^{\prime}$ orthogonal to the columns of the
      design matrix $M$. Similarly,
      the data compression we suggest in Section~\ref{sec:lincom} is
      equivalent to marginalising over vectors that lie in the
      subspace orthogonal to the column space of $H$ with uniform priors.
      By considering the data in the basis orthogonal to the column space of $H$
      to be nuisance parameters with uniform priors, the resulting likelihood of
      the compressed generalised residuals becomes independent of the value of
      the data in the orthogonal complement (we have not found another prior
      with the same property). This interpretation of
      data compression in terms of marginalisation assures us that we are not
      introducing any biases or unwanted systematics in our analysis. The
      difference with the marginalisation over the timing model is that we do
      not marginalise over physical nuisance parameters; we are throwing away
      information. The data compression matrix $H$ as constructed in
      Section~\ref{sec:lincom} guarantees that we throw away as little
      information about the signal amplitude $a$ as possible. This is not true
      for the other parameters this signal may also depend on: optimal
      sensitivity to those parameters possibly requires a different basis,
      construction of which is subject of ongoing follow-up research. We ignore
      this issue in the rest of this exploratory work, and assume that
      sensitivity to the signal amplitude is sufficient for our purposes.

      The interpretation of data compression in terms of marginalising over
      non-physical parameters assures us that the likelihood of the
      compressed data in Equation~(\ref{eq:compressedlikelihood}) is also valid
      if we do not provide good estimators for $\Sigma_w$ and $S$. We may be
      throwing away more information than we thought if our estimators are not
      accurate, but we do not introduce any bias or systematics in our
      likelihood. It is therefore not imperative to be thorough in the estimates
      of the signal and the noise; a reasonable guess may be sufficient for
      practical purposes.

\begin{figure}
	\includegraphics[width=0.5\textwidth]{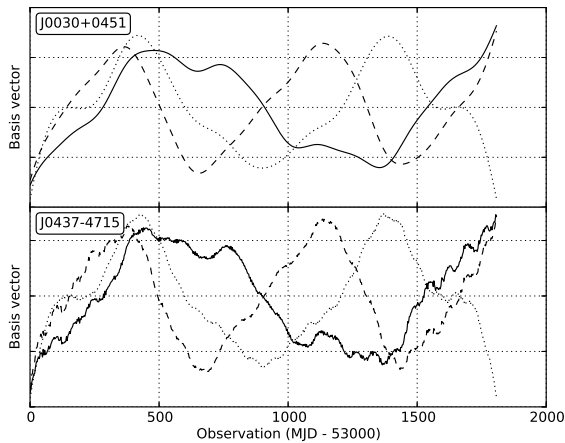}
	\caption{The first three compressed basis vectors of J0030+0451 and
	  J0437-4715 of IPTA Mock Data Challenge open $1$. These basis vectors are the the first
	  three columns of the matrix $GH$. The basis vectors are normalised, so
	  we have ignored the scaling on the y-axis.
	  The basis vectors of J0437-4715 have
	  more high-frequency structure due to the fact that J0437-4715 is in a
	  binary.}
	\label{fig:compressedbases}
      \end{figure}

      It is instructive to inspect the compressed basis vectors $GH$ for highly
      compressible signals. We choose the IPTA Mock Data Challenge as an
      example, since the GWB signal strongly dominates the noise at the lowest
      frequencies in these datasets. In Figure~\ref{fig:compressedbases} we
      present the first three compressed basis vectors for J0030+0451 and
      J0437-4715 of Mock Data Challenge open $1$. We
      observe that, roughly, the first basis vector corresponds to a third-order
      polynomial: start negative, then ascend to a maximum, descend to a minimum,
      and finally end positive. The other two basis vectors display a similar
      behaviour with the order of the polynomial equal to the order of the basis
      vector $+2$. Note that the zeroth, first, and second order are missing due
      to the removal of quadratics in our marginalisation over the timing model
      parameters.

      We note that the compressed basis vectors for both pulsars in
      Figure~\ref{fig:compressedbases} are similar, except that those of
      J0437-4715 display more high-frequency behaviour. This is because
      J0437-4715 resides in a binary, and the timing-model therefore includes
      parameters for binary motion.

    \subsection{Compressibility: how far can we go?} \label{sec:compressibility}
      A natural question that arises in data compression is how much we can
      compress the data without losing a significant amount of information. To
      answer this question, we consider the fidelity as a function of the number
      of generalised residuals in the dataset. For compressible datasets we expect the
      fidelity to stay close to one, only to drop for high compression rates.
      One possible measure of compressibility, which we use in our application
      of the ABC-method, is the maximum compression for which the fidelity stays
      above $0.99$. This maximum compression depends on the signal amplitude and
      power spectral density compared to that of the noise.

\begin{figure}
	\includegraphics[width=0.5\textwidth]{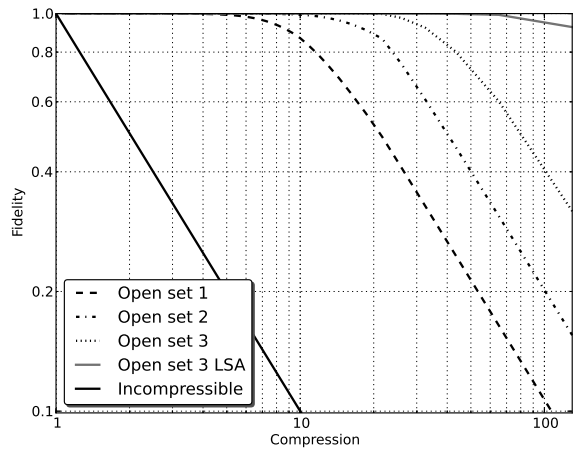}
	\caption{The fidelity of the GWB signal as a function of the
	  compression
	  for the pulsar J0030+0451 of all IPTA open data challenge
	  sets. Open dataset 3 contains more redundant information than the
	  other two sets because the GWB is smaller, and the GWB signal is
	  therefore buried under the noise at a larger portion of the spectrum.
	  Open dataset 2 contains more redundant information than open dataset
	  $1$
	  because the errorbars for J0030+0451 observations were set higher in
	  open dataset $2$ than in open dataset $1$. We have also plotted the
	  fidelity of the 
	  low-signal approximation (open $3$ LSA) for open dataset $3$, and the
	  fidelity of an incompressible signal (incompressible).}
	\label{fig:fidelityopen}
      \end{figure}

      As an example, we plot the fidelity of the mock data of J0030+0451
      from the open Mock Data Challenge versus the compression in
      Figure~\ref{fig:fidelityopen}, where the signal of interest is the
      gravitational-wave background. In these datasets the noise is white. Open
      dataset 3 does formally contain some extra (mildly) red noise which we
      do include in these plots, but the level of red noise is so low that it
      is negligible in practice. Because the signal is of such a different
      spectral shape than the noise, data compression is very efficient. In such
      a case, the higher the noise level compared to the signal, the more
      compressible the dataset is.

      In Figure~\ref{fig:fidelityopen} we plot the fidelity for
      open dataset $3$ in the low-signal limit (LSA).
      Data compression is most efficient in the low-signal limit. As a
      comparison we show the fidelity for an incompressible signal in
      Figure~\ref{fig:fidelityopen} as well. This corresponds to the high-signal
      limit. We see that an increase in the compression results in an equal
      decrease in the fidelity.

    \subsection{Compressing realistic datasets: a prescription}
      For realistic datasets we generally do not know the details of the signal
      and the noise. The noise typically has to be characterised from the data,
      and we may not even be certain of the presence of a signal of interest.
      Since the fidelity depends on estimates of the signal and the noise, it is
      not clear how far exactly we can compress the dataset without losing
      information from the signal. Here, we therefore recommend a conservative
      approach when preparing the ABC-method.

      The $a\lambda_i$ in the denominator of Equation~(\ref{eq:fidelity})
      represents the signal relative to the noise. The larger it is relative to
      $1$, the less likely we will discard that generalised residual. Therefore,
      if we are sure not to overestimate the noise, and if we are sure not to
      underestimate the signal, the compression fidelity will not be
      overestimated. Specifically, we recommend to calculate the fidelity as
      follows:\newline
      1) Construct the noise covariance estimate $\Sigma_w$ such that it only
      consists of the TOA uncertainties.\newline
      2) Choose a suitable spectral form for the signal of interest. For
      example: this consists of fixing the spectral index $\gamma = 13/3$ for
      the GWB.\newline
      3) Use the estimates of vHL
      \citep[Equation (22) \& (24) of][]{vanhaasteren2012} to estimate the
      signal amplitude. For a GWB signal, this is:
      \begin{equation}
	\label{eq:vhlampest}
	\sigma_{\rm GWB} = 1.37 \times 10^{-9}\left(\frac{A_h}{10^{-15}}\right)
	\left(\frac{T}{\rm yr}\right)^{\frac{5}{3}},
      \end{equation}
      where $T$ is the duration of the experiment, $A_h$ is the dimensionless
      GWB amplitude, and $\sigma_{\rm GWB}$ is the rms residual due to the GWB
      in the data. For other power spectral densities a similar calculation to
      vHL is required.

      By completely ignoring other effects like red spin noise in these
      estimates, we are ensured that we do not throw away more information than
      we should. Indeed, more noise in this calculation would mean a higher
      compression. This conservative approach is therefore also guaranteed to
      work in the presence of (strong) red noise.

      We note that this approach can overestimate the fidelity if the TOA
      uncertainties have been overestimated, or when the shape of the signal
      power spectral density has been estimated incorrectly with, for instance,
      an incorrect spectral index.  The TOA uncertainties depend on complex
      details of the data reduction pipeline prior to the formation of the TOAs
      and of the cross-correlation of the pulse profile with a template
      \citep{Taylor1992}. However, underestimation of the TOA uncertainty is
      uncommon in practice. How to choose a suitable basis to be sensitive to
      the spectral index is a subject of ongoing follow-up research. Here we
      assume we know the spectral index of the signal of interest.
      
      In the open Mock Data Challenge, shown in Figure~\ref{fig:fidelityopen},
      high compressions of over $50$ still yield a fidelity close to
      $\mathscr{F}=1$ in the low-signal limit. Since realistic datasets are
      expected to be in the low-signal limit - we have not detected a GWB yet -
      we expect high compressions in realistic datasets to be possible as well.
      However, realistic datasets can have far more TOAs per pulsar than
      the $130$ TOAs per pulsar in the Mock Data Challenge. Since in the
      low-signal limit only a few generalised residuals per pulsar is enough to reach
      $\mathscr{F} \geq 0.99$, we expect very high compressions, possibly up to
      $c = 1000$ depending on the size of the dataset, to be realistic for
      initial PTA applications.

  \section{Linear data compression in practice} \label{sec:practice}
    Although the raw data of pulsar observations can be quite voluminous, the
    pulsar time of arrival data files are typically several kilobytes in size.
    Because it seems quite unlikely that data volume at this stage of the analysis
    is ever going to be a problem, the only reason to resort to data
    compression is because it can greatly accelerate the analysis of pulsar
    timing data. In this section, we discuss the computational costs of
    evaluating the likelihood function with the ABC-method, 
    and we present some computational shortcuts. A straightforward application
    is a Bayesian analysis
    (e.g. vHLML), but other analysis methods described in the time domain are
    expected to see an equally large acceleration (e.g. D12).
    Special attention is given to power-law signals, for which we present a
    convenient approximation of the compressed covariance matrix, thereby
    maximising the effectiveness of data compression.

    \subsection{Computational demand}
      The computational demand of Equation~(\ref{eq:marginalisedlikelihoodnew})
      scales as $n^3$ (vHL) due to the inversion
      operation of an $(n \times n)$ matrix. With linear data
      compression, we have decreased the size of the inversion matrix, which
      will therefore also decrease the computational demands. The computational
      demand of the inversion in Equation~(\ref{eq:compressedlikelihood}) scales
      as $l^3$. Depending on the compression, this $l^3$ operation may or
      may not be the computational bottleneck. For large enough compression
      factors, the computational bottleneck will either be the computation of
      $C$ ($n^2$ operation), or
      the multiplication $H^{T}CH$ ($ln^2$ operation). In the case of an array
      of pulsars, the matrix $H$ will be block-diagonal if the data compression
      has been done per individual pulsar. Then,
      the computation of $H^{T}CH$ can be accelerated with a
      factor of the number of pulsars by block-wise multiplication (vHL).

      In this assessment of computational demand, we have neglected the
      construction of the data compression matrix $H$. A computationally
      expensive singular value decomposition of a full covariance matrix is
      required for this. However, this only needs to be done once: we do not
      change the compressed basis during subsequent likelihood evaluations, even
      if we vary the noise/signal parameters during a Markov Chain Monte Carlo
      simulation. Since the compressed basis can be calculated for each pulsar
      individually, we therefore do not expect the construction the data
      compression matrix $H$ to be a computational bottleneck in the foreseeable
      future.

    \subsection{Testing the acceleration of the likelihood} \label{sec:profiling}
      We test the performance of the ABC-method on the IPTA Mock Data
      Challenge: all
      challenges consist of $130$ observations per pulsar, with $36$ pulsars. Our
      likelihood contains the following deterministic and stochastic signal
      contributions:\newline
      1) the {\rm Tempo2} \citep{Hobbs2006} timing-model parameters\newline
      2) error bars for every TOA\newline
      3) power-law red timing noise for every pulsar\newline
      4) a correlated GWB\newline
      Evaluation of the likelihood of Equation~(\ref{eq:marginalisedlikelihood})
      took on average $38.3$ seconds\footnote{All computations in this work are
      performed on a single workstation, code linked with an Automatically Tuned
      Linear Algebra System (ATLAS) library that came with the GNU/Linux
      distribution.}, where most of that time comes from inverting the full
      covariance matrix.

\begin{figure}
	\includegraphics[width=0.5\textwidth]{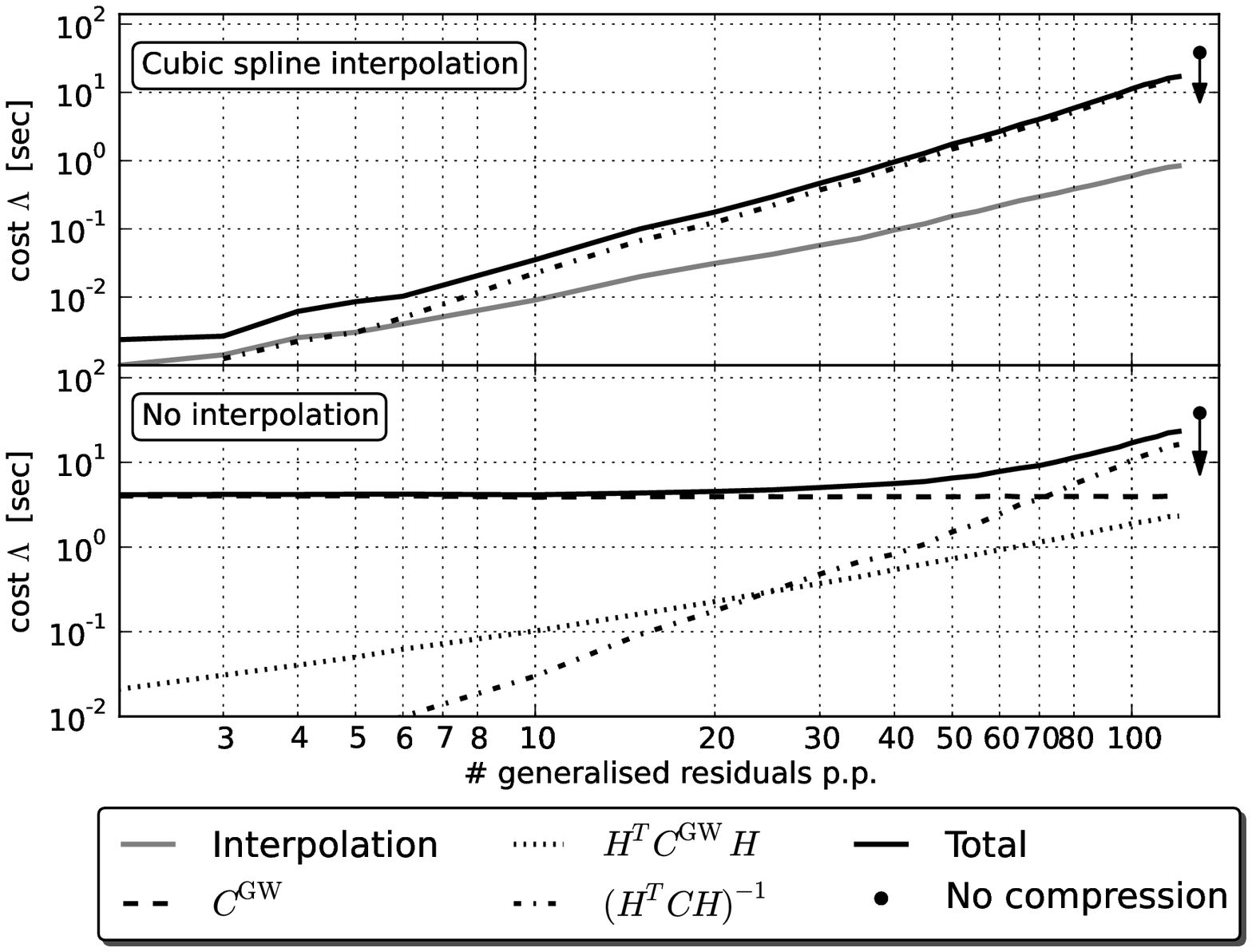}
	\caption{The computational cost of the dominating terms in the
	  compressed likelihood, as a function of the number of compressed
	  generalised residuals 
	  per pulsar. An array of $36$ pulsars was used, with $130$
	  observations per pulsar. The dominating terms are:\newline
	  1) $C^{\rm GW}$, (dashed line), the evaluation of the $(n\times n)$
	  elements of the covariance matrix of the GWB. Only present in the
	  lower panel.\newline
	  2) $H^{T}C^{\rm GW}H$, (dotted line), the matrix multiplication to
	  obtain the compressed covariance matrix. Only present in the lower
	  panel.\newline
	  3) Interpolation, (gray solid line), the construction of the $(l\times
	  l)$ compressed covariance matrix $H^{T}CH$ by cubic spline
	  interpolation. Only present in the upper panel.\newline
	  4) $(H^{T}CH)^{-1}$, (dash-dotted line), inversion of the compressed
	  covariance matrix.\newline
	  The total computational cost is shown as a solid line, and the
	  uncompressed likelihood of Equation~(\ref{eq:marginalisedlikelihood})
	  is shown as an upper limit at $130$ generalised residuals per pulsar.\newline
	  In the lower panel, these terms are evaluated for the compressed
	  likelihood of Equation~(\ref{eq:compressedlikelihood}), without any
	  computational shortcuts. For high compression factors (low number of
	  compressed generalised residuals), the evaluation of $C^{\rm GW}$ is
	  dominant, which means that further compression does not buy one more
	  computational time.\newline
	  In the upper panel the compressed likelihood is evaluated, where the
	  cubic spline interpolation method of
	  Section~\ref{sec:cubicspline} is used to evaluate $C^{\rm GW}$. In
	  this case, the inversion $(H^{T}CH)^{-1}$ is always the dominant term,
	  and data compression is most efficient. Note how the line for
	  $(H^{T}CH)^{-1}$ is (nearly) identical in both panels.}
	\label{fig:profiling}
      \end{figure}

      We compare the efficiency of Equation~(\ref{eq:marginalisedlikelihood}) to
      that of the data compression likelihood of
      Equation~(\ref{eq:compressedlikelihood}), where the latter equation
      becomes Equation~(\ref{eq:marginalisedlikelihoodnew}) when the compression
      is $1$. In the evaluation of the compressed likelihood, three terms take
      up the majority of the computational cost:\newline
      1) $C^{\rm GW}$, the evaluation of the $(n\times n)$ elements of the
      covariance matrix of the GWB.\newline
      2) $H^{T}C^{\rm GW}H$, the matrix multiplication to obtain the compressed
      covariance matrix.\newline
      3) $(H^{T}CH)^{-1}$, inversion of the compressed covariance matrix.\newline
      All other operations are negligible compared to these three. In
      Figure~\ref{fig:profiling} we present the computational cost of these
      three terms, together with the sum of the three, in the bottom panel. The
      uncompressed likelihood is given as a single point. We see that the
      inversion of the compressed covariance matrix is the dominant term for low
      compression factors: if roughly $70$ or more generalised residuals per
      pulsar are kept. For higher compression factors, the evaluation of
      $C^{\rm GW}$ is the most time-consuming part of the evaluation of the
      likelihood. Because this is an $n^{2}$ operation that does not depend
      on the compression, compressing the data to less than $50$ generalised
      residuals per pulsar does not gain us any computational efficiency in this
      configuration.

    \subsection{Signals with unknown amplitude} \label{sec:unamp}
      As explained in the previous section, in the case where a dataset
      is highly compressible, the computational bottleneck becomes
      evaluating $C$, which contains $C^{\rm GW}$ in the example of
      Section~\ref{sec:profiling}, at each step of the likelihood function. If
      we label the contributions to the compressed covariance matrix as
      $H^{T}\Sigma_{i}H$, then in some cases it is possible to greatly
      accelerate the evaluation of $H^{T}\Sigma_{i}H$. The simplest type
      of stochastic signal is the type where the power spectral density shape is
      known completely, but the amplitude $N_{i}$ is an unknown model parameter.
      Examples of signals of this type include the stochastic behaviour due to
      TOA uncertainties (with an unknown scaling, or "EFAC", parameter), pulse
      phase jitter \citep[e.g. ][]{Cordes2010}, or a GWB with a known spectral
      index.  For these types of signal we can evaluate $H^{T}\Sigma_{i}H$ just
      once for unit amplitude, and store this in memory. Then, each time we need
      to evaluate the likelihood function, we can multiply this stored matrix
      with the amplitude $N_{i}$ to obtain the compressed covariance matrix
      without having to re-calculate such matrices every time. Especially when
      $l \ll n$, this greatly reduces the time necessary to evaluate
      $H^{T}\Sigma_{i}H$.

    \subsection{Power-law signals} \label{sec:cubicspline}
      Most stochastic signal models have more free parameters than only an
      amplitude, and the acceleration method of Section~\ref{sec:unamp} is not
      applicable. In this section we
      present a practical solution for signals with two free parameters: an
      amplitude, and some other parameter. We focus only on signals with a
      power-law power spectral density, but we expect that the method is
      also appropriate for other signals with a parametrised power spectral
      density.

      Power-law signals are used in various ways in pulsar timing, both as a model
      for noise sources \citep[i.e. red spin noise][]{Cordes2010, Shannon2010},
      and as signal sources \citep[i.e. the istotropic background of gravitational
      waves][]{Phinney2001, vanhaasteren2009}. We use the following
      definition for the power spectral density of a power-law signal:
      \begin{equation}
	S(f) = a^{2}\left(\frac{1}{1 \rm{yr}^{-1}}\right)\left(\frac{f}{1
	\rm{yr}^{-1}}\right)^{-\gamma},
	\label{eq:powerlawspectraldensity}
      \end{equation}
      where $f$ is the signal frequency, $a$ is the signal amplitude, and
      $\gamma$ is the spectral index that describes the steepness of the
      spectrum. The rms in the timing residuals of such a signal is given by:
      $\sigma^{2}_{\rm rms} = \int_{0}^{\infty}\! \rm{d}f\, S(f)$.
      Because this is an unphysical power spectrum that diverges at
      the low frequencies, in practice a third parameter is used to describe a
      power-law signal that represents a lowest frequency $f_L$ below which the
      signal is assumed to be zero. The reduced data $\vec{\delta t}$ and
      therefore also the compressed data $\vec{x}$ are not affected by $f_L$
      \citep[vHL;][]{Blandford1984, Lee2012}.

      For highly compressed data, the compressed covariance matrix $H^{T}CH$
      contains far less elements than $C$: the number of unique elements for
      this matrix is $l(l+1)/2$. For a single pulsar power-law noise covariance
      matrix this is typically only of the order of a hundred elements, depending
      on the number of observations and the compression. We propose to use an
      interpolation approximation for
      each element of the matrix $H^{T}CH$ as a function of $\gamma$, with
      $1 < \gamma < 7$. The elements of the covariance matrix diverge at both
      ends of the interval. In the case of a single pulsar, this means we have
      $l(l+1)/2$ functions on the interval $1 < \gamma < 7$ that we want to
      write an interpolation approximation for.
      We choose a cubic spline interpolation method for this, where the domain
      of the function is divided in sub-intervals in which the function is
      approximated by a third-order polynomial. We construct all polynomials
      such that their values and derivatives match at the edges. The only free
      parameter in this approach is the number of cubics used in total. This
      number needs to be tuned for performance.

      \begin{figure}
	\includegraphics[width=0.5\textwidth]{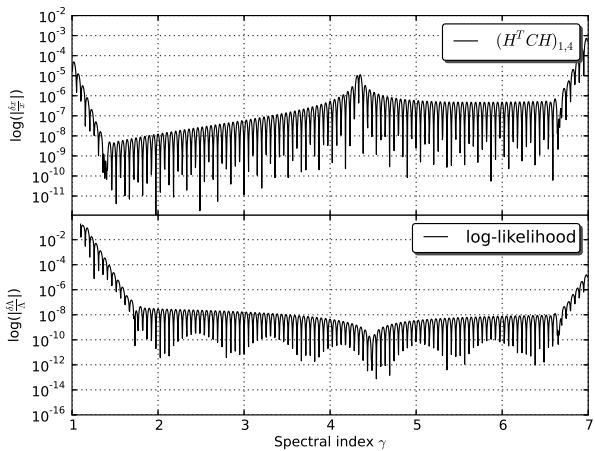}
	\caption{The likelihood and the covariance matrix $H^{T}CH$ as a
	  function of $\gamma$.
	  For the pulsar J0030+0451, with data as in IPTA Mock Data
	  Challenge open $2$, we
	  used the interpolation technique of
	  Section~\ref{sec:cubicspline} to approximate the elements of the
	  compressed covariance matrix $H^{T}CH$. In the upper panel, we have
	  plotted $log|\delta x / x|$ of element $x$ (row 1, column 4) of
	  the compressed covariance matrix as a function of the spectral index
	  $\gamma$. Here $x$ is the true value of the element of $H^{T}CH$, and
	  $\delta x$ is the
	  difference between the true value of $x$, and the interpolated value.
	  This plot looked similar for all elements. In the bottom panel we have
	  plotted the corresponding quantity for the log-likelihood: $log|\delta
	  \Lambda / \Lambda|$, with $\Lambda$ the log-likelihood, and $\delta
	  \Lambda$ the difference between the true and interpolated value. We
	  initialised the cubic spline interpolation with $100$ points, evenly
	  distributed on the interval $(1,7)$. We see that the discrepancy
	  between the interpolated and the true values grows steeply near the
	  boundaries of the interval. At the boundaries, the elements of the
	  compressed covariance matrix diverge.}
	\label{fig:interpolelements}
      \end{figure}

      In Figure~\ref{fig:interpolelements} we show the difference between the
      true value and the interpolated value of an arbitrary element of $H^{T}CH$
      as a function of $\gamma$ for J0030+0451 of Mock Data Challenge open $2$. These results are
      typical; we find a similar plot for every element, where the difference
      between the true value and the interpolated value always inflates near the
      boundaries of the interval. We also show the difference between the
      accompanying log-likelihood $\Lambda$ as a function of $\gamma$ for the
      same dataset. Here we also see that the difference inflates near the
      boundaries. The precision of the interpolation depends on the number of
      cubic splines used in the interpolation. For lower numbers of splines in
      the approximation, we saw the accuracy quickly decrease near the
      boundaries. This caused the compressed covariance matrix to become
      non-positive definite or singular close to the boundaries. In our
      simulations, $100$ equally spaced cubic splines was enough on a slightly
      reduced interval $1.09 < \gamma < 6.91$ to not run into numerical issues.

      The cubic spline interpolation removes the necessity to calculate the
      total covariance matrix $C$. In the top panel of
      Figure~\ref{fig:profiling} we present the computational cost of the
      computationally dominant terms in the compressed likelihood, in the case
      where we use cubic spline interpolation for the elements of $H^{T}CH$. The
      computationally dominant term is the inversion $(H^{T}CH)^{-1}$ for the
      whole range of possible compressions, which means that data compression is
      maximally efficient. We almost reached full capacity of Random Access
      Memory of our
      workstation for very low compressions. For large datasets
      with an incompressible signal, this may cause problems for the cubic
      spline interpolation method. However, for current applications, we don't
      believe this to be an issue. For the Mock Data Challenge, the total
      typical speedup at $99\%$ fidelity is almost three orders of magnitude.

    \subsection{Tests on the IPTA Mock Data Challenge} \label{sec:mdc}
\begin{figure}
	\includegraphics[width=0.5\textwidth]{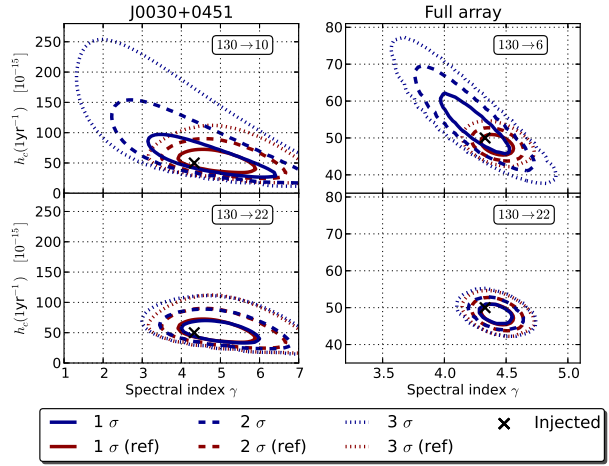}
	\caption{The compressed likelihood as a function of the GWB amplitude
	  and spectral index $\gamma$ for IPTA Mock Data Challenge open one. No parameters are
	  numerically marginalised over. The timing model parameters are
	  analytically marginalised over as part of the data compression. On the
	  left panel the likelihood is plotted for only pulsar J0030+0451, with
	  compression to $10$ generalised residuals (top), and compression to
	  $22$ generalised residuals 
	  (bottom). On the right the likelihood is plotted for the full
	  array of pulsars, with compression to $6$ generalised residuals per pulsar
	  (top), and compression to $22$ generalised residuals per pulsar
	  (bottom). In each panel, the blue lines represent the credible regions
	  of the compressed likelihood, the red lines, labelled "ref", represent
	  the reference credible regions of the uncompressed likelihood of
	  Equation~(\ref{eq:marginalisedlikelihoodnew}). The contours represent
	  the $1\sigma$ ($68\%$), $2\sigma$ ($95\%$), and $3\sigma$ ($99.7\%$)
	  credible regions. The injected values are marked with an 'x'.}
	\label{fig:gammasensitivity}
      \end{figure}

      We test the ABC-method with the cubic spline interpolation technique
      on the open Mock Data Challenge. We present the results here of Mock Data
      Challenge open $1$ because the
      noise level was the same for all pulsars in that challenge. That makes it
      easier to compare the results we see here with the fidelity levels of
      Figure~\ref{fig:fidelityopen}: they are approximately the same for all
      pulsars. In Figure~\ref{fig:fidelityopen} we see that for a compression
      of $6$, we start to approach $\mathscr{F} \approx 0.99$. This
      corresponds to $22$ compressed generalised residuals per pulsar. In
      Figure~\ref{fig:gammasensitivity} we present the likelihood credible
      regions for
      Mock Data Challenge open $1$ both for the full array of pulsars and for pulsar
      J0030+0451, with different compression levels. We see that with $22$
      generalised residuals
      per pulsar, the compressed likelihood is practically equal to the
      uncompressed likelihood, as predicted by Figure~\ref{fig:fidelityopen}.
      With less than $22$ generalised residuals per pulsar, the likelihood
      credible regions are broader, with significant covariance between the GWB
      amplitude and the spectral index. This covariance may partially be a
      result of the compressed basis being optimal only for the injected value
      of the spectral index $\gamma=4.33$; this dependence is the subject of
      follow up work.

      The results of this section hold for all three of the open Mock Data
      Challenge datasets:
      when the fidelity $\mathscr{F} \geq 0.99$, the likelihood credible regions
      where almost indistinguishable from the uncompressed likelihood credible
      regions. With a compression such that the fidelity is significantly less
      than that, the credible regions were broader, with a covariance between
      the amplitude and spectral index.

  \section{Conclusions} \label{sec:conclusions}
    We investigate the acceleration of the analysis of pulsar timing data by
    compressing the data with a linear transformation, without losing a
    significant amount of information of a particular stochastic signal of
    interest: the ABC-method.
    In this formalism, marginalisation over the timing-model parameters is
    equivalent to lossless linear data compression. We show that when
    the stochastic signal of interest has a significantly different spectrum
    than the noise, the data is highly compressible. The ABC-method is most
    efficient in the low-signal limit, where the signal is buried under the
    noise over most of the frequency range. Data compression is not
    possible in the strong-signal limit, where the signal dominates the noise in
    the whole frequency range. The likelihood function of the compressed signal
    is computationally more efficient, and unbiased.

    We introduce the concepts of compression and compression fidelity, where the
    compression is the total number of timing residuals divided by the number of
    generalised timing residuals 
    that are kept in the compression, and the fidelity is a measure of the
    amount of information about the signal of interest that is kept in the
    compression. For the IPTA Mock Data Challenge, we show that the compression
    is of the order of $10$, at a fidelity $\mathscr{F} = 0.99$, if one is
    interested in the isotropic stochastic background of gravitational waves.

    When applied to highly compressible datasets, computational shortcuts are
    required to optimally accelerate the evaluation of the compressed
    likelihood. We present an practical method based on cubic spline interpolation
    of the compressed covariance matrix. When this interpolation approximation
    is used, the total acceleration of the evaluation of the compressed
    likelihood is $c^3$, with $c$ the compression.
    We test the cubic spline interpolation
    method, and conclude that it works well for the purposes of the IPTA Mock
    Data Challenge. The total acceleration is about three orders of magnitude
    for a compression of $10$, with results almost identical to an analysis
    without the ABC-method.

    The ABC-method can be readily applied to realistic datasets,
    without any adjustments.  Realistic datasets of current Pulsar Timing Arrays
    are expected to reside in the low-signal approximation: no stochastic
    gravitational-wave background has been detected as of yet. Therefore, a high
    compression factor of several hundred is realistic for such datasets, which
    yields a total acceleration of over six orders of magnitude. We expect
    linear data compression to become one of the key solutions for the issues
    related to computational cost in pulsar timing array data analysis.

  \section*{Acknowledgements}
    The author thanks K.J. Lee for some insightful discussions, and Alberto
    Sesana for providing the inspiration for this work. Yuri Levin and Michele
    Vallisneri are thanked for their extensive comments on this manuscript. The
    research in this paper is conducted as part of the efforts of the European
    Pulsar Timing Array (EPTA).


  \bibliographystyle{mn2e.bst}

\begin{thebibliography}{}

\bibitem[\protect\citeauthoryear{{Begelman}, {Blandford} \& {Rees}}{{Begelman}
  et~al.}{1980}]{Begelman1980}
{Begelman} M.~C.,  {Blandford} R.~D.,    {Rees} M.~J.,  1980, Nature, 287, 307

\bibitem[\protect\citeauthoryear{{Blandford}, {Romani} \&
  {Narayan}}{{Blandford} et~al.}{1984}]{Blandford1984}
{Blandford} R.,  {Romani} R.~W.,    {Narayan} R.,  1984, Journal of
  Astrophysics and Astronomy, 5, 369

\bibitem[\protect\citeauthoryear{{Cordes} \& {Shannon}}{{Cordes} \&
  {Shannon}}{2010}]{Cordes2010}
{Cordes} J.~M.,  {Shannon} R.~M.,  2010, ArXiv e-prints

\bibitem[\protect\citeauthoryear{{Damour} \& {Vilenkin}}{{Damour} \&
  {Vilenkin}}{2005}]{Damour2005}
{Damour} T.,  {Vilenkin} A.,  2005, \prd, 71, 063510

\bibitem[\protect\citeauthoryear{{Demorest}, {Ferdman}, {Gonzalez}, {Nice},
  {Ransom}, {Stairs}, {Arzoumanian} \& {Brazier}}{{Demorest}
  et~al.}{2012}]{Demorest2012}
{Demorest} P.~B.,  {Ferdman} R.~D.,  {Gonzalez} M.~E.,  {Nice} D.,  {Ransom}
  S.,  {Stairs} I.~H.,  {Arzoumanian} Z.,    {Brazier} A.,  2012, ArXiv
  e-prints

\bibitem[\protect\citeauthoryear{Detweiler}{Detweiler}{1979}]{Detweiler1979}
Detweiler S.,  1979, \apj, 234, 1100

\bibitem[\protect\citeauthoryear{{Edwards}, {Hobbs} \& {Manchester}}{{Edwards}
  et~al.}{2006}]{Edwards2006}
{Edwards} R.~T.,  {Hobbs} G.~B.,    {Manchester} R.~N.,  2006, \mnras, 372,
  1549

\bibitem[\protect\citeauthoryear{Estabrook \& Wahlquist}{Estabrook \&
  Wahlquist}{1975}]{Estabrook1975}
Estabrook F.,  Wahlquist H.,  1975, \grg, 6, 439

\bibitem[\protect\citeauthoryear{Foster \& Backer}{Foster \&
  Backer}{1990}]{Foster1990}
Foster R.,  Backer D.,  1990, \apj, 361, 300

\bibitem[\protect\citeauthoryear{{Grishchuk}}{{Grishchuk}}{2005}]{Grishchuk200%
5}
{Grishchuk} L.~P.,  2005, Uspekhi Fizicheskikh Nauk, 48, 1235

\bibitem[\protect\citeauthoryear{{Hobbs}, {Edwards} \& {Manchester}}{{Hobbs}
  et~al.}{2006}]{Hobbs2006}
{Hobbs} G.~B.,  {Edwards} R.~T.,    {Manchester} R.~N.,  2006, \mnras, 369, 655

\bibitem[\protect\citeauthoryear{Jaffe \& Backer}{Jaffe \&
  Backer}{2003}]{Jaffe2003}
Jaffe A.,  Backer D.,  2003, \apj, 583, 616

\bibitem[\protect\citeauthoryear{{Kramer}, {Stairs}, {Manchester},
  {McLaughlin}, {Lyne}, {Ferdman}, {Burgay}, {Lorimer}, {Possenti}, {D'Amico},
  {Sarkissian}, {Hobbs}, {Reynolds}, {Freire} \& {Camilo}}{{Kramer}
  et~al.}{2006}]{Kramer2006}
{Kramer} M.,  {Stairs} I.~H.,  {Manchester} R.~N.,  {McLaughlin} M.~A.,  {Lyne}
  A.~G.,  {Ferdman} R.~D.,  {Burgay} M.,  {Lorimer} D.~R.,  {Possenti} A.,
  {D'Amico} N.,  {Sarkissian} J.~M.,  {Hobbs} G.~B.,  {Reynolds} J.~E.,
  {Freire} P.~C.~C.,    {Camilo} F.,  2006, Science, 314, 97

\bibitem[\protect\citeauthoryear{{Lee}, {Bassa}, {Janssen}, {Karuppusamy},
  {Kramer}, {Smits} \& {Stappers}}{{Lee} et~al.}{2012}]{Lee2012}
{Lee} K.~J.,  {Bassa} C.~G.,  {Janssen} G.~H.,  {Karuppusamy} R.,  {Kramer} M.,
   {Smits} R.,    {Stappers} B.~W.,  2012, \mnras, 423, 2642

\bibitem[\protect\citeauthoryear{{{\"O}lmez}, {Mandic} \&
  {Siemens}}{{{\"O}lmez} et~al.}{2010}]{Olmez2010}
{{\"O}lmez} S.,  {Mandic} V.,    {Siemens} X.,  2010, \prd, 81, 104028

\bibitem[\protect\citeauthoryear{{Phinney}}{{Phinney}}{2001}]{Phinney2001}
{Phinney} E.~S.,  2001, ArXiv Astrophysics e-prints

\bibitem[\protect\citeauthoryear{Press, Teukolsky, Vetterling \&
  Flannery}{Press et~al.}{1992}]{Press1992}
Press W.,  Teukolsky S.,  Vetterling W.,    Flannery B.,  1992, {Numerical
  Recipes in C}, 2nd edn.
Cambridge University Press, Cambridge, UK

\bibitem[\protect\citeauthoryear{{Sanidas}, {Battye} \& {Stappers}}{{Sanidas}
  et~al.}{2012}]{Sanidas2012}
{Sanidas} S.~A.,  {Battye} R.~A.,    {Stappers} B.~W.,  2012, \prd, 85, 122003

\bibitem[\protect\citeauthoryear{{Sazhin}}{{Sazhin}}{1978}]{Sazhin1978}
{Sazhin} M.~V.,  1978, Soviet Astronomy, 22, 36

\bibitem[\protect\citeauthoryear{{Sesana}, {Vecchio} \& {Colacino}}{{Sesana}
  et~al.}{2008}]{Sesana2008}
{Sesana} A.,  {Vecchio} A.,    {Colacino} C.~N.,  2008, \mnras, 390, 192

\bibitem[\protect\citeauthoryear{{Shannon} \& {Cordes}}{{Shannon} \&
  {Cordes}}{2010}]{Shannon2010}
{Shannon} R.~M.,  {Cordes} J.~M.,  2010, \apj, 725, 1607

\bibitem[\protect\citeauthoryear{{Taylor}}{{Taylor}}{1992}]{Taylor1992}
{Taylor} J.~H.,  1992, Philosophical Transactions of the Royal Society of
  London, 341, 117-134 (1992), 341, 117

\bibitem[\protect\citeauthoryear{{Taylor} \& {Weisberg}}{{Taylor} \&
  {Weisberg}}{1982}]{Taylor1982}
{Taylor} J.~H.,  {Weisberg} J.~M.,  1982, \apj, 253, 908

\bibitem[\protect\citeauthoryear{{Vallisneri}}{{Vallisneri}}{2008}]{Vallisneri%
2008}
{Vallisneri} M.,  2008, \prd, 77, 042001

\bibitem[\protect\citeauthoryear{{van Haasteren} \& {Levin}}{{van Haasteren} \&
  {Levin}}{2012}]{vanhaasteren2012}
{van Haasteren} R.,  {Levin} Y.,  2012, \mnras, p.~88

\bibitem[\protect\citeauthoryear{{van Haasteren}, {Levin}, {Janssen},
  {Lazaridis}, {Kramer}, {Stappers}, {Desvignes}, {Purver} \& {Lyne}}{{van
  Haasteren} et~al.}{2011}]{vanhaasteren2011}
{van Haasteren} R.,  {Levin} Y.,  {Janssen} G.~H.,  {Lazaridis} K.,  {Kramer}
  M.,  {Stappers} B.~W.,  {Desvignes} G.,  {Purver} M.~B.,    {Lyne} A.~G.,
  2011, \mnras, 414, 3117

\bibitem[\protect\citeauthoryear{{van Haasteren}, {Levin}, {McDonald} \&
  {Lu}}{{van Haasteren} et~al.}{2009}]{vanhaasteren2009}
{van Haasteren} R.,  {Levin} Y.,  {McDonald} P.,    {Lu} T.,  2009, \mnras,
  395, 1005

\bibitem[\protect\citeauthoryear{Wade \& Wade}{Wade \& Wade}{1994}]{Wade1994}
Wade G.,  Wade J.~G.,  1994, Signal Coding \& Processing, 2nd edn.
Cambridge University Press, New York, NY, USA

\bibitem[\protect\citeauthoryear{Wyithe \& Loeb}{Wyithe \&
  Loeb}{2003}]{Wyithe2003}
Wyithe J.,  Loeb A.,  2003, \apj, 595, 614

\end{thebibliography}

 

\end{document}